\documentclass[11pt]{article}
\usepackage[utf8]{inputenc}
\usepackage[T2A]{fontenc}
\usepackage[english]{babel}
\usepackage{graphicx}

\begin{document}
\author{I.A.~Kachaev, VES experiment}
\title{Modified S-wave $\pi\pi$ scattering amplitude for multiparticle PWA}
\date{\today}


\maketitle

\begin{abstract}
Suggested by Au, Morgan, Pennington (AMP) $S$-wave isospin $I=0$ $\pi\pi$, $KK$
scattering amplitude is good enough to describe experimental data for the moment.
Still it has two disadvantages for use in multiparticle 
partial wave analysis (PWA), namely sharp drop at the $KK$ threshold and
unreasonable behavior at $M(\pi\pi) > 1.6\,GeV/c^2$. The drop is not seen in
multiparticle systems.

We suggest the modified AMP amplitude, mAMP, for the only aim, namely to describe
the broad part of $S$-wave $\pi\pi\,\to\,\pi\pi$ scattering in the wide $M(\pi\pi)$
range in multiparticle PWA. The mAMP amplitude describes threshold
behavior of the $\pi\pi\,\to\,\pi\pi$ scattering and the wide structure at
$M\sim 1400\,MeV/c^2$ reasonably well.
It is assumed that narrow objects $f_0(980)$, $f_0(1500)$ are included in PWA
separately. The amplitude does not describe $\pi\pi\,\to\,KK$ scattering.
The mAMP amplitude is purely phenomenological.
\end{abstract}
\centerline{(to be submitted to Physics of Atomic Nuclei)}

\section{Introduction}

To perform multiparticle PWA we need to know two-particle amplitudes, the
\emph{isobars} in the PWA terminology.
For $\pi\pi$ system in $P$, $D$, $F$ waves two-particle amplitudes are mostly
described by known resonances $\rho(770)$, $f_2(1270)$, $\rho_3(1690)$ et al.
Still in the $\pi\pi$ $S$-wave the situation is more complicated.
In addition to narrow objects $f_0(980)$, $f_0(1500)$ there exists broad
structure including at least threshold peculiarities, wide $f_0(500)$, $f_0(1370)$
and probably some other objects.





Scattering amplitude in the $\pi\pi$ $S$-wave was investigated in details
in AMP article~\cite{AMP}. This article is concentrated on the construction
of the scattering amplitude in the channels $\pi\pi$, $KK$ 
with proper analyticity and unitarity.
The best experimental data known at this time are used
on input. The amplitude is presented in the form appropriate for
later use.



When AMP parametrization of $\pi\pi$ scattering was used in three-particle PWA 
of $\pi^-\pi^-\pi^+$ system the following problems are found. The AMP 
amplitude describes $\pi\pi$ $S$-wave in a whole, in the full range of 
$M(\pi\pi)$. This leads to both physical and technical restrictions.

From the physical point of view amplitudes and phases of resonances
in AMP amplitude are fixed once and forever. The authors of \cite{AMP}
used so called $P$-vector approach to describe production of the $\pi\pi$
system.


In the AMP amplitude there is a zero at $M(\pi\pi) \approx 1\,GeV/c^2$
due to $f_0(980)$ object and the threshold in the $KK$ channel.
If this zero is not seen in the multiparticle system it should be compensated
by the pole in $P$-vector. It is impossible in the multiparticle PWA
if in it a connection of $\pi\pi$ system with production channel is
described by a set of coupling coefficients, as it is done in the
Illinois PWA program~\cite{Hansen}.


From the technical point of view the AMP amplitude is parametrized via a set
of poles and background polynomial of 4th degree. It is known that polynomial
parametrization rapidly makes senseless outside of the range of definition.
Due to this fact AMP amplitude has unphysical maximum at 
$M(\pi\pi)\sim 1.6\,GeV/c^2$.

\section{The $K$-matrix method}


The AMP amplitude is constructed in $K$-matrix formalism \cite{Chung-KMX,Chung-KMX2}.
Formally the transition from initial to final state 
$S_{ij} = \left< j|S|i \right>$
is described by the unitary scattering operator, $SS^+ = S^+S = I$.


Still it is much simple to work with Hermitian matrices then with unitary.
Let us construct the Hermitian matrix $K$ which describes $S$ exactly.
After separation of no-interaction part one can define the transition operator $T$ as 
\begin{equation}
S = I + 2iT
\label{T-matrix}
\end{equation}
where $I$ is identity operator and factor $2i$ is introduced for convenience.
From unitarity of $S$ and the definition of $T$ we have
\begin{equation}
T - T^+ = 2i\,T^+T = 2i\,TT^+
\end{equation}
Introducing the inverse operator $T^{-1}$ we have
\begin{equation}
(T^+)^{-1} - T^{-1} = 2i\,I \quad\mbox{so}\quad 
(T^{-1} + i\,I)^+ = T^{-1} + i\,I
\label{K-herm}
\end{equation}
Now we can introduce operators $K$ and $M$, from (\ref{K-herm}) they are Hermitian
\begin{equation}
K^{-1} =  T^{-1} + i\,I \quad\mbox{and}\quad M = K^{-1}
\label{K-def}
\end{equation}
It is known that if time-reversal invariance is hold than $K$ and $M$ matrices are 
not only Hermitian but are also real and so symmetric. Explicit definition for $T$ is
\begin{equation}
T = K \left( I - i K \right)^{-1} =
         \left( I - i K \right)^{-1} K
\end{equation}

\smallskip
So defined transition amplitude $T$ is not Lorentz invariant. Let us construct
it in Lorentz invariant form. For decay of particle with mass $m$ to two particles
with masses $m_a$, $m_b$ phase space is $\rho = 2q/m$ where $q$ is breakup momentum
\begin{equation}
\rho = \sqrt{\left[ 1 - \left( {\frac{m_a+m_b}{m}} \right)^2 \right] 
             \left[ 1 - \left( {\frac{m_a-m_b}{m}} \right)^2 \right]}
\end{equation}
Phase space is normalized as $\rho \to 1$ at $m \to \infty$.
By definition phase space is considered a diagonal matrix, so in two channel case
it is
\begin{equation}
\rho = \left( \begin{array}{cc} \rho_1 & 0 \\ 0 & \rho_2 \end{array} \right)
\end{equation}
Lorentz invariant amplitude $\hat T$ is defined as
\begin{equation}
T_{ij} = \{ \rho_i^* \}^{1/2} \,{\hat T}_{ij}\, \{ \rho_j \}^{1/2} 
\label{T-matrix-hat}
\end{equation}
or in matrix form
\begin{equation}
T = \{ \rho^+ \}^{1/2} \,\hat T\, \{ \rho \}^{1/2} 
\end{equation}
Below the threshold of the channel its phase space become complex. Complex
conjugation in (\ref{T-matrix-hat}) is required for proper analytic continuation 
into the region below the threshold of some channels.

Let us note a significant difference between $T$ and $\hat T$. In the one channel case
\begin{equation}
S = e^{2i\delta}, \quad T = e^{i\delta}\sin\delta, \quad \hat T = \frac{1}{\rho}\, T
\end{equation}
where $\delta$ is phase shift. For any $\delta$ the amplitude $T$
is confined in the circle with the centre $(0,i/2)$ and the radius $1/2$.
The dependence $T(s)$ is named the Argand diagram. The amplitude $\hat T$
has other normalization and is used as Dalitz plot amplitude in PWA programs.
On the threshold of the reaction in the case of $S$-wave scattering
$T \to 0$ but $\hat T \to const$.

Now we can define Lorentz invariant matrix $\hat K$ as
\begin{equation}
K = \{ \rho^+ \}^{1/2} \,\hat K\, \{ \rho \}^{1/2}
\label{K-matrix-hat}
\end{equation}
If $\hat K$ is taken to be real and symmetric, then $K$ is Hermitian and $S$ unitary
even some $\rho$ becomes imaginary below the threshold of some channels.
From (\ref{K-def}) we have
\begin{equation}
{\hat K}^{-1} =  {\hat T}^{-1} + i\rho
\end{equation}
Explicit form of $\hat T$ is
\begin{equation}
\hat T = \hat K \left( I - i\rho \hat K \right)^{-1} =
         \left( I - i \hat K \rho \right)^{-1} \hat K
\end{equation}

\section{The original AMP amplitude}



In \cite{AMP} (3.18) $\hat K$ matrix (named below $K$ for consistency with \cite{AMP})
has been parametrized as sum of poles and polynomial background 
(this is so called $K$ solution)
\begin{equation}
K_{ij} = \frac{(s-s_0)}{4m^2_K} \sum_p \frac{f_i^p f_j^p}{(s_p-s)(s_p-s_0)}
+ \sum_{n=0} c_{ij}^n \left(\frac{s}{4m^2_K}-1\right)^n
\label{K-param}
\end{equation}
%
By definition $m_K = 1/2(m_{K+}+m_{K0})$. Here $s_0$ is known Adler zero near the
threshold of $\pi\pi$ system. We have found that amplitude calculated according
to this formula does not match fig.\,(5.3) in \cite{AMP}.
If we use parameters stated in the article proper formula which corresponds
to this figure is
\begin{equation}
K_{ij} = \frac{(s-s_0)}{4m^2_K} \left[ \sum_p \frac{f_i^p f_j^p}{(s_p-s)(s_p-s_0)}
+ \sum_{n=0} c_{ij}^n \left(\frac{s}{4m^2_K}-1\right)^n \right]
\label{K-param-b}
\end{equation}
We believe that it is formula (\ref{K-param-b}) that describes the 
original $K$ solution of AMP amplitude.

For the $\hat M$ matrix in \cite{AMP} (3.20)
there exists the following parametrization (so called $M$ solution)
\begin{equation}
M_{ij} = \frac{a_{ij}}{(s-s_0)} + \sum_p \frac{{f}^p_i {f}^p_j}{(s_p-s)}
+ \sum_{n=0} {c}_{ij}^{n} \left[\frac{s}{4m^2_K}-1\right]^n
\label{M-param}
\end{equation}
%
Again, we believe that there is a mistake in the sign in \cite{AMP} here
and the proper formula for $M$ solution is
\begin{equation}
M_{ij} = \frac{a_{ij}}{(s-s_0)} - \sum_p \frac{{f}^p_i {f}^p_j}{(s_p-s)}
+ \sum_{n=0} {c}_{ij}^{n} \left[\frac{s}{4m^2_K}-1\right]^n
\label{M-param-b}
\end{equation}
Note that for $M$ solution parameters of $T$ matrix tends to zero at $s\to\infty$
while for $K$ solution they tends to unitary limit. The reason of this behavior is
that outside of its scope polynomial background tends to infinity.
We consider the behavior of $M$ solution more appropriate for our aims and use it.
Parameters of original $M$ solution from \cite{AMP} are listed in table \ref{table-coef}
\begin{table}[!hb]
\begin{center}
\begin{tabular}{|c|c|c|c|c|c|c|}
\hline
$s_0$  & $s_1$ & $f^1_1$ & $f^1_2$ & $a_{11}$ & $a_{12} $ & $a_{22}$ \\
\hline
-0.0074 & 0.9828 & 0.1968 & -0.0154 & 0.1131 & 0.0150 & -0.3216 \\
\hline
\end{tabular}

~\\[4pt]

\begin{tabular}{|c|c|c|c|c|}
\cline{1-5}
$c^0_{11}$ & $c^1_{11}$ & $c^2_{11}$ & $c^3_{11}$ & $c^4_{11}$ \\
\cline{1-5}
0.0337 & -0.3185 & -0.0942 & -0.5927 & 0.1957 \\
\cline{1-5}
$c^0_{12}$ & $c^1_{12}$ & $c^2_{12}$ & $c^3_{12}$ & $c^4_{12}$ \\
\cline{1-5}
-0.2826 & 0.0918 & 0.1669 & -0.2082 & -0.1386 \\
\cline{1-5}
$c^0_{22}$ & $c^1_{22}$ & $c^2_{22}$ & $c^3_{22}$ & $c^4_{22}$ \\
\cline{1-5}
0.3010 & -0.5140 & 0.1176  & 0.5204 & -0.3977 \\
\cline{1-5}
\end{tabular}
\end{center}
\caption{Table of coefficients of original AMP amplitude}
\label{table-coef}
\end{table}

\begin{table}[!htb]
\begin{center}
\begin{tabular}{|c|c|c|c|c|c|c|}
\hline
$s_0$  & $s_1$ & $f^1_1$ & $f^1_2$ & $a_{11}$ & $a_{12} $ & $a_{22}$ \\
\hline
-0.0074 & 0.9828 & 0 & 0 & 0.1131 & 0 & -0.3216 \\
\hline
\end{tabular}

~\\[4pt]

\begin{tabular}{|c|c|c|c|c|}
\cline{1-5}
$c^0_{11}$ & $c^1_{11}$ & $c^2_{11}$ & $c^3_{11}$ & $c^4_{11}$ \\
\cline{1-5}
0.0337 & -0.3185 & -0.0942 & -0.5927 & 0 \\
\cline{1-5}
$c^0_{12}$ & $c^1_{12}$ & $c^2_{12}$ & $c^3_{12}$ & $c^4_{12}$ \\
\cline{1-5}
0        & 0          & 0           & 0           & ~~0~~ \\
\cline{1-5}
$c^0_{22}$ & $c^1_{22}$ & $c^2_{22}$ & $c^3_{22}$ & $c^4_{22}$ \\
\cline{1-5}
0.3010 & -0.5140 & 0.1176  & 0.5204 & ~~0~~ \\
\cline{1-5}
\end{tabular}
\end{center}
\caption{Table of coefficients for modified amplitude mAMP}
\label{table-coef-m}
\end{table}

\section{Our modifications}


Our aim is to construct the scattering amplitude $\pi\pi\to\pi\pi$
which is near to the original AMP amplitude, is smooth in the region
$M(\pi\pi) \approx 1\,GeV/c^2$ and smoothly tends to zero at 
$M(\pi\pi) > 1.5\,GeV/c^2$.

We require the proper behavior at $M(\pi\pi) \approx 1\,GeV/c^2$ by setting
to zero the connection with $KK$ channel and the connection with $f_0(980)$
pole. To suppress an unphysical shoulder in the original amplitude at
$M(\pi\pi)\sim 1.7 \,GeV/c^2$ we set to zero coefficients at the 4th degree
of the background polynomial.

\begin{figure}[t!]
\centerline{\includegraphics[width=\textwidth]{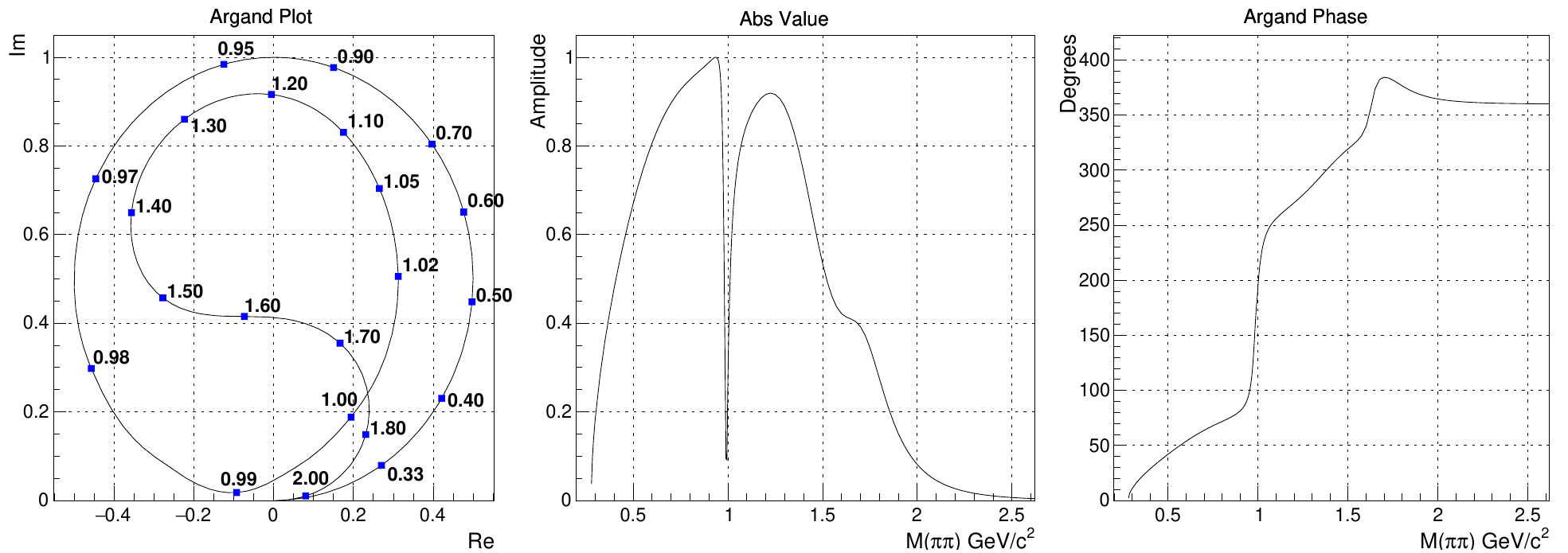}}
\caption{Original AMP amplitude}
\label{fig-AMP}
\end{figure}

\begin{figure}[b!]
\centerline{\includegraphics[width=\textwidth]{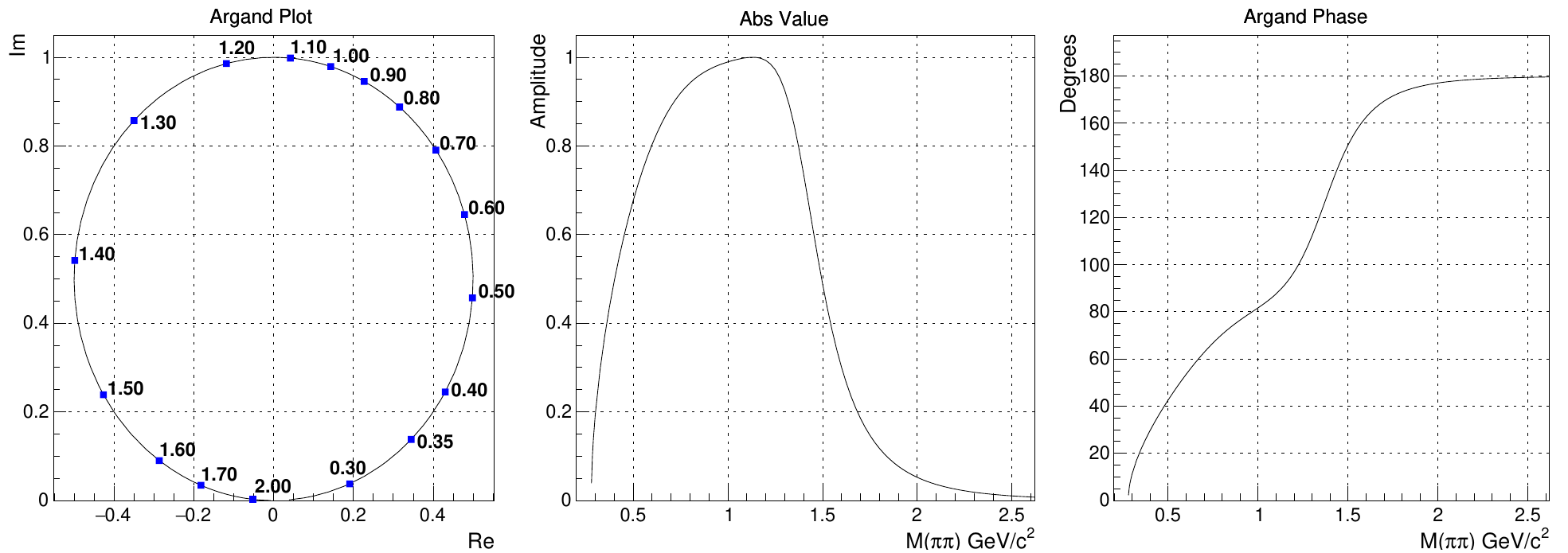}}
\caption{Modified AMP amplitude}
\label{fig-AMP-M}
\end{figure}

The parameters of the modified amplitude are given in the table 
\ref{table-coef-m}. The original amplitude is shown in the figure \ref{fig-AMP},
the modified one in the figure \ref{fig-AMP-M}. The figures show the values,
from left to right: Argand plot of the matrix element $T_{11}$, namely 
$\pi\pi\to\pi\pi$, its absolute value and Argand phase.


\section{Results}



The amplitude mAMP is constructed, suitable to describe $\pi\pi\to\pi\pi$ 
scattering in $S$ wave with $I=0$ in multiparticle PWA. It is not suitable
to describe $KK$ channel.

It is unitary, smooth in the broad range of $M(\pi\pi)$, is near to AMP $M$ solution
for $M(\pi\pi) < 1.6\ GeV/c^2$ except at $M(\pi\pi) \sim 1\,GeV/c^2$,
smoothly tends to zero at $M(\pi\pi)\to\infty$.
It describes the threshold behavior of $\pi\pi$ scattering and the broad 
structure at $M\sim 1400\,MeV/c^2$ according to the data known at the time
of writing of \cite{AMP}. Narrow resonances $f_0(980)$, $f_0(1500)$ intentionally
are not described, they should be entered into the analysis separately.




The amplitude is built according to formula (\ref{M-param-b}) with coefficients
from table \ref{table-coef-m}.
This amplitude was used in \cite{pippm,MESON2018} and the other publications of
the VES group and also in \cite{COMPASS-3pi}.
An early version of the amplitude was used in \cite{Chung-3pi}.

The author thanks the members of the VES group for initiating this work, 
useful discussions and use of this amplitude in the analysis.


\end{document}